\documentclass[11pt]{article}
\usepackage{amsmath,amssymb}
\usepackage[dvips]{epsfig}
\newtheorem{theorem}{Theorem}[section]
\newtheorem{lemma}[theorem]{Lemma}
\newtheorem{corollary}[theorem]{Corollary}

\newtheorem{remark}[theorem]{Remark}
\protect \hoffset -1 true cm \voffset -1.5 true cm \textwidth 15cm
\newcommand{\res[1]}{\operatornamewithlimits{Res}_{#1}}
\newcommand{\vu}{\boldsymbol{u}}

\newcommand{\ra}{\rightarrow}
\newcommand{\lb}{\lambda}

\newcommand{\C}{\mathcal{C}}

\begin{document}

\title{Reciprocal transformations and flat metrics on Hurwitz spaces}

\author {Simonetta Abenda \\
Dipartimento di Matematica e CIRAM \\
Universit\`a degli Studi di Bologna, Italy \\
{\footnotesize  abenda@ciram.unibo.it} \\
and \\
Tamara Grava \\
SISSA, Via Beirut 9, Trieste, Italy  \\
{\footnotesize  grava@sissa.it}} \maketitle
\begin{abstract}
We consider hydrodynamic systems which possess a local Hamiltonian
structure. To such a system there are
also associated an infinite number of nonlocal Hamiltonian
structures. We give necessary and
sufficient conditions so that, after a nonlinear transformation of
the independent variables, the reciprocal system still possesses a
local Hamiltonian structure. We show
that, under our hypotheses, bi--hamiltonicity is preserved by the
reciprocal transformation. Finally we apply such results to
reciprocal systems of genus $g$ Whitham-KdV modulation equations.
\end{abstract}

\section{Introduction}

Systems of hydrodynamic type that admit Riemann invariants are a
class of quasilinear evolutionary PDEs of the form
\begin{equation}
\label{sistini}
u^i_t= v^i(\vu)u^i_x, \quad i=1,\dots, n,
\end{equation}
where $\vu=(u^1,\dots,u^n)$ (see e.g.
\cite{DN1,DN2,DN3,Du2,Olver,T,Tsa}). The systems (\ref{sistini})
admit a local Hamiltonian structure if there exists a
nondegenerate flat diagonal metric $g_{ii} (\vu)(du^i)^2$ solution
to \cite{Tsa}
\begin{equation}\label{tsa}
\partial_j \ln \sqrt{g_{ii} (\vu)} = \frac{\partial_j
v^i(\vu)}{v^j(\vu)-v^i(\vu)}.
\end{equation}
The corresponding Hamiltonian operator
\begin{equation}
\label{DN}
J^{ij} (\vu) = g^{ii} (\
vu) \left(\delta^i_j \frac{d}{dx}-
\Gamma^{j}_{ik} (\vu) u^k_x\right),\quad g^{ii}=1/g_{ii},
\end{equation}
with  $\Gamma^i_{jk} (\vu)$ are the Christoffel symbols of the
metric $g_{ii}(\vu)$, defines a Poisson bracket on functionals
\[
\{A,B\}=\int\displaystyle \frac{\delta A}{\delta
u^i(x)}J^{ij}\frac{\delta A}{\delta u^k(x)}dx.
\]
Such local Hamiltonian structures were introduced by
Dubrovin-Novikov \cite{DN1} and we refer to (\ref{DN}) as DN
Hamiltonian structures. The Hamiltonian form of the equations
(\ref{sistini})  is
\begin{equation}
\label{hamilton}
u^i_t=\{u^i, H\}=J^{ij} (\vu)\partial_j h(\vu)= v^i(\vu)u^i_x, \quad
i=1,\dots, n,
\end{equation}
where $H=\int h(\vu)dx$ is the Hamiltonian. The system
(\ref{hamilton}) possesses an infinite number of conservation laws
and commuting flows and it is integrable through the generalized
hodograph transform \cite{Tsa}. The formula (\ref{tsa}) is crucial
for the integrability property of diagonal Hamiltonian systems: if
one interprets it as an overdetermined system on $n$ unknown
functions $v^i(\vu)$, ($g_{ii}(\vu)$ given), one can generate for
any other  solution $w^i(\vu)$, a symmetry $u^i_{\tau}=
w^i(\vu)u^i_x$ of (\ref{hamilton}), namely
$(u^i_t)_{\tau}=(u^i_{\tau})_{t}$. One can prove the completeness
property of this class of symmetries which implies integrability
\cite{Tsa}. For any symmetry $w^i(\vu)$ of the Hamiltonian system
(\ref{DN}), one can define the metric
$\tilde{g}_{ii}(\vu)=(w^i(\vu))^2g_{ii}(\vu)$,  which is still
flat and it is related to the metric $g_{ii}(\vu)$ by a Combescure
transformation.

From a differential geometric point of view, a non-degenerate flat
diagonal metric is equivalent to giving an orthogonal coordinate
system on a flat space. Locally this coordinate system is
parameterized by $n(n-1)/2$ functions of two variables. The
problem of determining orthogonal coordinate systems dates back to
the 19th century (see \cite{Zakharov} and references therein). In
the case $n=2$ the problem has been solved classically, while it
is still open for $n\geq 3$. Zakharov \cite{Zakharov} showed that
the problem can be solved by the dressing method.

All non--trivial examples of flat metrics have been obtained in
the framework of the theory of Frobenius manifolds \cite{Du2,Du0},
when the metric is of Egorov type. We recall that a metric
$g_{ii}(\vu)$ is Egorov, if its rotation coefficients
\[
\ \beta_{ij} (\vu) \equiv \frac{\partial_i  \sqrt{g_{jj} (\vu)}}{
\sqrt{g_{ii} (\vu)}}, \quad i\not = j,
\]
are symmetric, namely $\beta_{ij}(\vu)=\beta_{ji}(\vu)$.

\smallskip

In this paper we address the problem of finding nontrivial
examples of flat metric not of Egorov type, applying reciprocal
transformations to the Hamiltonian structures of DN systems.

Indeed any DN system also possesses an infinite number of nonlocal
Hamiltonian structures (see \cite{FM,Fer,MN,Ma}), since equation
(\ref{tsa}) defines $g_{ii}(\vu)$ up to a multiple
$g_{ii}(\vu)/f^i(u^i)$, where $f^i(u^i)$ is an arbitrary function
of $u^i$. Although the metric $g_{ii}(\vu)$ may happen to be flat
for a particular choice of $f^i(u^i)$, it will not be flat in
general. In particular, if the metric $g_{ii} (\vu)$ is of
constant Riemannian curvature $c$ or conformally flat with
curvature tensor
\begin{equation}\label{Rij}
R^{ij}_{ij} (\vu) = w^i(\vu) + w^j(\vu), \quad i\not =
j,
\end{equation} where the $w^i(\vu)$ satisfy
(\ref{tsa}), then the Hamiltonian operator associated to
(\ref{sistini}) is nonlocal and takes the special form
\begin{equation}
\begin{split}
\label{MF}
J^{ij} (\vu)& = g^{ii} (\vu) \left(\delta^i_j
\frac{d}{dx}- \Gamma^{j}_{ik} (\vu) u^k_x\right)+c u^i_x \left( \frac{d}{dx}
\right)^{-1}  u^j_x,\\
J^{ij} (\vu) &= g^{ii} (\vu) \left(\delta^i_j \frac{d}{dx}-
\Gamma^{j}_{ik} (\vu) u^k_x\right)+ w^i (\vu) u^i_x \left(
\frac{d}{dx} \right)^{-1}  u^j_x + u^i_x \left( \frac{d}{dx}
\right)^{-1}  w^j (\vu)u^j_x,
\end{split}
\end{equation}
respectively. The first operator was introduced by Ferapontov and
Mokhov \cite{FM}, while the second one by Ferapontov\cite{Fer}.

\smallskip

Reciprocal transformations are a class of transformations of the
independent variables and were introduced in gas dynamic
\cite{RS}. Assuming that the DN hydrodynamic system
(\ref{sistini}) admits conservation laws
\[
B(\vu)_t=A(\vu)_x,\quad N(\vu)_t=M(\vu)_x
\]
with $B(\vu)M(\vu)-A(\vu)N(\vu)\neq 0$, then we can perform a
change of the independent variables $(x,t)\rightarrow
(\hat{x}(x,t,\vu),\hat{t}(x,t,\vu)$ by the relations
\begin{equation*}
d{\hat x} = B (\vu)dx + A(\vu)dt,\quad\quad d{\hat t} = N(\vu)dx
+M(\vu)dt.
\end{equation*}
Then the reciprocal system
\begin{equation*}
u^{i}_{{\hat t}}
=\frac{B(\vu)v^i(\vu)-A(\vu)}{M(\vu)-N(\vu)v^i(\vu)}u^{i}_{{\hat
x}}={\hat v}^i (\vu) u^{i}_{{\hat x}},
\end{equation*}
is clearly a system of hydrodynamic type.

Since reciprocal transformations send conservation laws to
conservation laws, it is natural to investigate their effect on
the corresponding Hamiltonian structure.
The reciprocal metric is given by the formula
\[
{\hat g}_{ii}(\vu)=
\left(\frac{M(\vu)-N(\vu)v^i(\vu)}{B(\vu)M(\vu)-A(\vu)N(\vu)}
\right)^2{ g}_{ii}(\vu)
\]
and clearly if ${ g}_{ii}(\vu)$ is flat ${\hat g}_{ii}(\vu)$ is in general  not flat.
Linear reciprocal
transformations, namely when $B(\vu), A(\vu), M(\vu)$ and $N(\vu)$
are constants, preserve flatness of the metric and locality of DN
Hamiltonian operators (see Tsarev \cite{Tsa} and Pavlov
\cite{Pav}). In the case of nonlinear reciprocal transformations
\cite{FP}, Ferapontov  and Pavlov have proven that the reciprocal
to a flat metric is, in general, conformally flat. Moreover,
Ferapontov~\cite{Fer} gave necessary and sufficient condition for
the reciprocal to a flat metric to be a constant curvature in case
the reciprocal transformation is a linear combination of Casimirs,
momentum and the Hamiltonian density.

In a recent paper \cite{AG}, we have proven that  the
Camassa--Holm (CH) modulation equations admit a local
bi-hamiltonian structure of DN type and the corresponding flat
metrics  are reciprocal to the constant curvature and conformally
flat metric of the Korteweg-de Vries (KdV) modulation equations.
It is remarkable that none of the metrics of CH Hamiltonian
structures  are of Egorov type. For the above reasons, we are
interested in a systematic investigation on the conditions under
which the reciprocal to a (non)-flat metric is flat.

\smallskip

In this manuscript we work out necessary and sufficient conditions
for the reciprocal metric to be flat, when the initial metric is
either flat or constant curvature or conformally flat. The
necessary and sufficient conditions for reciprocal flat metrics of
sections 4-6 can be applied to search new examples of flat metrics
on Hurwitz spaces. As a by-product we obtain non-trivial examples
of flat metrics which are non-Egorov on the moduli space of
hyperelliptic Riemann surfaces.

Finally,  supposing that the initial system is bi-hamiltonian,
namely it possesses two compatible Hamiltonian operators (see
\cite{Magri,Du2,Du0,DLZ,Fer2,Fordy,Mok1,Mok01}), we give
sufficient conditions such that the reciprocal hydrodynamic system
is bi-hamiltonian as well. We recall that bi--hamiltonicity is
preserved by linear transformations~\cite{XZ}.

The plan of the paper is as follows. In section 2 we set the
notation and we compute the reciprocal Riemannian curvature tensor
and the reciprocal Hamiltonian structure for any metric associated
to the initial system. In section 3 we give sufficient conditions
for the bi--hamiltonicity of the reciprocal to a bi--hamiltonian
system when the transformation is nonlinear. In section 4 and 5 we
consider the case of reciprocal transformations in $x$
(respectively $t$) and we present the complete set of necessary
and sufficient conditions for a reciprocal metric to be flat, when
the initial metric is either flat or of constant curvature or
conformally flat (respectively flat). In section 7 we consider
reciprocal transformations of both variables $x$ and $t$ and we
give sufficient conditions for a reciprocal metric to be flat,
when the initial metric is either flat or of constant curvature or
conformally flat. All of the necessary and sufficient conditions
in sections 5--7 are expressed in Riemann invariants of the
initial system and are compatible with the results in~\cite{Fer},
where applicable.

Finally, in section 8, we give examples of flat reciprocal metrics
on the moduli space of hyperelliptic Riemann surfaces. In
particular, we relate  by a reciprocal transformation the genus
$g$ Whitham--KdV hierarchy to the genus $g$ Whitham--Camassa--Holm
hierarchy.

\section{The reciprocal Hamiltonian structure}
In the following, we consider a DN Hamiltonian hydrodynamic system
in Riemann invariants as in (\ref{DN})
\begin{equation}\label{sisdia}
u^i_t =v^i(\vu)u^i_x.
\end{equation}
Let $g^{ii} (\vu)$ be a non--degenerate metric such that for
convenient $f^i(u^i)$, $i=1,\dots, n$, $g^{ii} (\vu)f^i(u^i)$ is a
flat metric associated to the local Hamiltonian operator of the
system (\ref{sisdia}). Let $H_i (\vu)$, $\beta_{ij}(\vu)$ and ${
\Gamma}^{i}_{jk}(\vu)$  be respectively the Lam\'e coefficients
the rotation coefficients and the Christoffel symbol of a diagonal
non-degenerate metric $g_{ii} (\vu)$ associated to
(\ref{sistini}),
\[
H_i (\vu) = \sqrt{g_{ii} (\vu)},\quad\quad \beta_{ij} (\vu) =
\frac{\partial_i H_j (\vu)}{H_i(\vu)}, \quad i\not = j,
\]
\[
{ \Gamma}^{i}_{jk}(\vu)= \frac{1}{2} g^{im} (\vu) \left(
\frac{\partial  g_{mk}(\vu)}{\partial u^j} + \frac{\partial
g_{mj}(\vu)}{\partial u^k} -\frac{\partial g_{kj}(\vu)}{\partial
u^m} \right),
\]
then the nonzero  elements of the Riemannian curvature tensor are
\begin{equation}\label{curvijk}
\begin{array}{l}
\displaystyle R^{ij}_{ik} (\vu) = -\frac{\partial_k
\beta_{ij}(\vu)-\beta_{ik}(\vu)\beta_{kj}(\vu)}{H_i (\vu)
H_j(\vu)}\equiv 0,\quad\quad i\not = j\neq k\\
\displaystyle R^{ik}_{ik} (\vu) = -\frac{\Delta_{ik}(\vu)}{H_i
(\vu) H_k(\vu)} \equiv \sum_{(l)} \epsilon^{l} w^{i}_{(l)} (\vu)
w^{k}_{(l)} (\vu) ,\quad\quad i\not = k
\end{array}
\end{equation}
where $\epsilon^{l}=\pm 1$, $w^{i}_{(l)} (\vu)$ are affinors of
the metric and
\[
\Delta_{ik}(\vu)=\partial_i \beta_{ik} (\vu)+\partial_k
\beta_{ki}(\vu) +\sum_{m\not = i,k} \beta_{mi}
(\vu)\beta_{mk}(\vu),
\]
and the Hamiltonian operator associated to $g^{ii} (\vu)$ is of
nonlocal  type \cite{FM,Fer}
\begin{equation}\label{ham}
J^{ij} (\vu) =g^{ii} (\vu) \left(\delta^i_j \frac{d}{dx}-
\Gamma^{j}_{ik} (\vu) u^k_x\right)+\sum_{l} \epsilon^{(l)}
w^i_{(l)} (\vu) u^i_x \left( \frac{d}{dx} \right)^{-1} w^j_{(l)}
(\vu) u^j_x.
\end{equation}
If $g^{ii} (\vu)$ is either flat or constant curvature or
conformally flat, $R^{ij}_{ij} (\vu)$ is either zero or constant
or as in (\ref{Rij}) and $J^{ij} (\vu)$ takes the form (\ref{DN})
or (\ref{MF}), respectively.

Given conservation laws
\[
B(\vu)_t=A(\vu)_x,\quad N(\vu)_t=M(\vu)_x
\]
for the system (\ref{sisdia}), a reciprocal transformation of the
independent variables $x,t$ is defined by \cite{RS}
\begin{equation}\label{xt}
d{\hat x} = B (\vu)dx + A(\vu)dt,\quad\quad d{\hat t} = N(\vu)dx
+M(\vu)dt.
\end{equation}
Then the reciprocal system
\begin{equation}\label{sistfin}
u^{i}_{{\hat t}} = {\hat v}^i (\vu) u^{i}_{{\hat x}}
=\frac{B(\vu)v^i(\vu)-A(\vu)}{M(\vu)-N(\vu)v^i(\vu)}u^{i}_{{\hat
x}},
\end{equation}
is still Hamiltonian with ${\hat J}^{ij}(\vu)$ Hamiltonian
operator associated to the reciprocal metric
\begin{equation}\label{tramet}
{\hat g}_{ii}(\vu)=
\left(\frac{M(\vu)-N(\vu)v^i(\vu)}{B(\vu)M(\vu)-A(\vu)N(\vu)}
\right)^2{ g}_{ii}(\vu).
\end{equation}
Let ${\hat H}_i(\vu)$, ${\hat \beta}_{ij}(\vu)$, ${
\hat{\Gamma}}^{i}_{jk}(\vu)$ and ${\hat R}^{ij}_{km}(\vu)$, be
respectively, the Lam\'e coefficients the rotation coefficients
and the Christoffel symbol for the reciprocal metric ${\hat
g}_{ii}(\vu)$. In the following  we compute their expressions and
that of the operator $\hat{J}^{ij}$. In \cite{FP}, Ferapontov and
Pavlov have characterized the tensor of the reciprocal Riemannian
curvature and the reciprocal Hamiltonian structure when the
initial metric $g_{ii}(\vu)$ is flat. To simplify notations, we
drop the $\vu$ dependence in the lengthy formulas.

\begin{theorem}\label{theo3.1}
Let ${ g}^{ii} (\vu)$ be the contravariant diagonal metric as
above for the Hamiltonian system (\ref{sisdia}) with Riemannian
curvature tensor as in (\ref{curvijk}) or in (\ref{Rij}). Then,
for the contravariant reciprocal metric ${\hat
g}^{ii}(\vu)=1/{\hat g}_{ii}(\vu)$ defined in (\ref{tramet}), the
only possible non-zero components of the reciprocal Riemannian
curvature tensor are
\begin{equation}\label{tracurvd}
\begin{array}{rcl}
\displaystyle {\hat R}^{ik}_{ik} (\vu)&=&\displaystyle \frac{{
H}_i{ H}_k}{{\hat H}_i{\hat H}_k}{ R}^{ik}_{ik} -({ \nabla} B)^2+
\frac{{ H}_k}{{\hat H}_k}{ \nabla}^k{ \nabla}_k B+\frac{{
H}_i}{{\hat H}_i}{ \nabla}^i{ \nabla}_i B-{\hat
v}^k{\hat v}^i({ \nabla} N)^2\\
&&\\
\displaystyle &&\displaystyle +{\hat v}^k\frac{{ H}_i}{{\hat
H}_i}{ \nabla}^i{ \nabla}_i N+{\hat v}^i\frac{{ H}_k}{{\hat H}_k}{
\nabla}^k{ \nabla}_k N- ({\hat v}^k+{\hat v}^i)<{ \nabla} B,{
\nabla} N>,\quad i\not=k\\
\end{array}
\end{equation}
where
\[
<{ \nabla} B(\vu),{ \nabla} N(\vu)> = \sum_{m} { g}^{mm}(\vu)
\partial_m B (\vu)\,
\partial_m N(\vu),\]
\[
{ \nabla}^i{ \nabla}_i B(\vu) = { g}^{ii} (\vu)\left(
\partial_i^2 B (\vu)-\sum_{m} { \Gamma}^m_{ii}(\vu)\,
\partial_m B (\vu)\right),
\]
\[
{ \nabla}^i{ \nabla}_j B (\vu)= { g}^{ii} (\vu)\left(
\partial_i\partial_j B (\vu)-{ \Gamma}^i_{ij}(\vu)\,
\partial_i B(\vu) -{ \Gamma}^j_{ij}(\vu)\,
\partial_j B(\vu)\right).
\]
\end{theorem}

\medskip

{\sl Proof.} \hspace{.5 truecm} To compute the reciprocal
Riemannian curvature tensor, we first compute the reciprocal
rotation coefficients. Since the initial system is Hamiltonian,
the rotation coefficients of the initial metric $g^{ii} (\vu)$
satisfy $\displaystyle \beta_{ik} (\vu) =\frac{\partial_i H_k
(\vu)}{H_i(\vu)} = \frac{\partial_i (H_k (\vu)v^k(\vu)) }{
H_i(\vu) v^i(\vu)}$. Moreover $\displaystyle v_i
(\vu)=\frac{\partial_i M(\vu)}{\partial_i N(\vu)}
=\frac{\partial_i A(\vu)}{\partial_i B(\vu)}$. Using the above
expressions, it is straightforward to verify that the reciprocal
rotation coefficients satisfy
\[
\begin{array}{rl}
{\hat \beta}_{ik} (\vu) \,\equiv&\,\displaystyle \frac{\partial_i
{\hat H}_k(\vu) }{{\hat H}_i(\vu)}\\
=&\displaystyle { \beta}_{ik} -\frac{(M-Nv_k){ H}_k\partial_i B}{{
H}_i (BM-AN)}+\frac{(M-Nv_k){ H}_k\partial_i N
(A-Bv_i)}{(BM-AN)(M-Nv_i){ H}_i} + \frac{{ H}_k(v_i-v_k)\partial_i
N}{(M-Nv_i){ H}_i} \\
=&\displaystyle\,{ \beta}_{ik}(\vu)  - {\hat H}_k(\vu)
\frac{\partial_i B(\vu)}{{ H}_i(\vu) } - {\hat H}_k (\vu) {\hat
v}_k(\vu) \frac{\partial_i N(\vu)}{{ H}_i(\vu) }.
\end{array}
\]
For the diagonal metric ${\hat g}^{ii} (\vu)$ the only possibly
non zero elements of the Riemannian curvature tensor are ${\hat
R}^{ij}_{ik}(\vu)$, ($i\not =j\not=k\not =i$) and ${\hat
R}^{ij}_{ij}(\vu)$, ($i\not =j$). To prove ${\hat
R}^{ij}_{ik}(\vu)=0$, $(i\not= j\not = k\not=i)$, we use
\[
\begin{array}{l}
\displaystyle \partial_j {\hat \beta}_{ik} (\vu)- {\hat
\beta}_{ij}(\vu){\hat \beta}_{jk} (\vu) =\displaystyle\partial_j
\left( { \beta}_{ik} -{\hat H}_k \frac{\partial_i B}{{ H}_i}
-{\hat H}_k{\hat v}_k\frac{\partial_i N}{{ H}_i}\right)\\
\displaystyle -\left({ \beta}_{ij} -{\hat H}_j \frac{\partial_i
B}{{ H}_i} -{\hat H}_j{\hat v}_j\frac{\partial_i N}{{
H}_i}\right)\left( { \beta}_{jk} -{\hat H}_k \frac{\partial_j B}{{
H}_j}
-{\hat H}_k{\hat v}_k\frac{\partial_j N}{{ H}_j}\right) \\
=\displaystyle \partial_j { \beta}_{ik}-{ \beta}_{ij}{
\beta}_{jk}- {\hat H}_k{\hat v}_k{ H}_i\nabla^i \nabla_j N(\vu)
-{\hat H}_k{ H}_i\nabla^i \nabla_j B(\vu),
\end{array}
\]
that is ${\hat R}^{ij}_{ik}(\vu)=0$ if and only if ${ \nabla}^i{
\nabla}_j B (\vu)=0={ \nabla}^i{ \nabla}_j N (\vu)$, $(i\not =j)$.
Indeed, the Darboux equations
\[
\partial_j { \beta}_{ik}(\vu)-{
\beta}_{ij}(\vu){ \beta}_{jk}(\vu) =0
\]
are equivalent to the condition $R^{ij}_{ik} (\vu)=0$. By
hypothesis, $ {\tilde g}_{ii}(\vu)= g_{ii}(\vu) / f^i(u^i)$ is a
flat metric, then the Christoffel symbols of the two metrics
satisfy ${\tilde \Gamma}^i_{ij} (\vu) =\Gamma^i_{ij} (\vu)$,
($i\not = j$), so that $\nabla^i \nabla_j B(\vu)=0=\nabla^i
\nabla_j N(\vu)$, ($i\not=j$) and, finally, ${\hat
R}^{ij}_{ik}(\vu) =0$.

\medskip

To prove (\ref{tracurvd}), we set
\[
\displaystyle {\hat R}^{ik}_{ik} (\vu)\; = \; -\frac{ {\hat
\Delta}_{ik}(\vu) }{{\hat H}_k (\vu){\hat H}_i (\vu)},\quad (i\not
=k), \quad {\rm where} \quad {\hat \Delta}_{ik} (\vu) = \partial_i
{\hat \beta}_{ik}  +
\partial_k {\hat \beta}_{ki}  + \sum_{m\not = i,k}
{\hat \beta}_{mi} {\hat \beta}_{mk} .
\]
Then
\[
\begin{array}{c}
\displaystyle {\hat \Delta}_{ik} (\vu) =\displaystyle {
\Delta}_{ik}  - \frac{{\hat H}_{k} }{ { H}_i} \left(
\partial_i^2 B-\sum_{m} { \Gamma}^m_{ii}
\partial_m B  \right) -
\frac{{\hat H}_{i}}{ { H}_k} \left(
\partial_k^2 B-\sum_{m} { \Gamma}^m_{kk}
\partial_m B  \right) \\
\displaystyle \quad - \frac{{\hat H}_{k} {\hat v}_k }{{
H}_i}\left(
\partial_i^2 N-\sum_{m} { \Gamma}^m_{ii}
\partial_m N  \right)
- \frac{{\hat H}_{i} {\hat v}_i}{ { H}_k} \left(
\partial_k^2 N-\sum_{m} { \Gamma}^m_{kk}
\partial_m N  \right)\\
\displaystyle +\sum_{m} \frac{{\hat H}_i  {\hat H}_k }{{
H}_m^2}\left( (\partial_m B)^2 +  {\hat v}_i{\hat v}_k (\partial_m
N)^2 + ({\hat v}_i  +{\hat v}_k )\partial_m B\,
\partial_m N\right)\end{array}
\]
\[
\begin{array}{c}
=\displaystyle { \Delta}_{ik}  - {\hat H}_{k}  { H}_i { \nabla}^i{
\nabla}_i B- {\hat H}_{i} { H}_k { \nabla}^k{ \nabla}_k B
\displaystyle - {\hat H}_{k} {\hat v}_k { H}_i { \nabla}^i{
\nabla}_i N- {\hat H}_{i} {\hat v}_i { H}_k { \nabla}^k{
\nabla}_k N\\
\\
+\displaystyle {\hat H}_i  {\hat H}_k \left( ({ \nabla} B)^2 +
{\hat v}_i{\hat v}_k ({ \nabla} N)^2 + ({\hat v}_i  +{\hat v}_k
)<{ \nabla} B\, ,\, { \nabla} N>\right).
\end{array}
\]
from which (\ref{tracurvd}) immediately follows. \hspace{.3
truecm} $\square$

\medskip

We now compute the reciprocal affinors and the reciprocal
Hamiltonian operator of a hydrodynamic system (\ref{sisdia}) with
(nonlocal) Hamiltonian operator (\ref{MF}). To this aim, we
introduce the auxiliary flows
\begin{equation}\label{bnflow}
u^i_{\tau} = n^{i} (\vu) u^i_x = { J}^{ij} (\vu)
\partial_j N(\vu),
\quad\quad \displaystyle u^i_{\zeta} = b^{i} (\vu) u^i_x = {
J}^{ij} (\vu)
\partial_j B(\vu),
\end{equation}
\[
u^i_{t_{(l)}} = w^i_{(l)}(\vu) u^i_x = { J}^{ij} (\vu)
\partial_j H^{(l)}(\vu),\]
respectively, generated by the densities of conservation laws
associated to the reciprocal transformation (\ref{xt}), $B(\vu)$,
$N(\vu)$, and by the densities of conservation laws $H^{(l)}(\vu)$
associated to the affinors $w^i_{(l)}$ of the Riemannian curvature
tensor (\ref{Rij}). By construction, all the auxiliary flows
commute with (\ref{sistini}). Introducing the following closed
form
\begin{equation}\label{clofor}
\left\{ \begin{array}{l} \displaystyle d{\hat x} = B (\vu)dx +
A(\vu)dt+ P(\vu)d\tau + Q(\vu) d\zeta + \sum_{l} T^{(l)}
(\vu)dt_{(l)},
\\\displaystyle d{\hat t} = N(\vu)dx + M(\vu)dt+R(\vu)d\tau+
S(\vu)d\zeta + \sum_{l} Z^{(l)}(\vu)
dt_{(l)},\\
\displaystyle d{\hat \tau} = d\tau, \quad d{\hat \zeta} = d\zeta,
\quad d{\hat t}_{(l)} = dt_{(l)},
\end{array}\right.\end{equation}
where $P(\vu)$, $S(\vu)$, $Q(\vu)$, $R(\vu)$, $T^{(l)}(\vu)$,
$Z^{(l)}(\vu)$ are defined up to additive constants, we have
\begin{equation}\label{clofl}
\begin{array}{l}
\displaystyle v^i(\vu) = \frac{\partial_i A(\vu)}{\partial_i
B(\vu)}= \frac{\partial_i M(\vu)}{\partial_i N(\vu)}, \quad\quad
w^i_{(l)}(\vu) = \frac{\partial_i T^{(l)} (\vu)}{\partial_i
B(\vu)}= \frac{\partial_i Z^{(l)}(\vu)}{\partial_i N(\vu)},\\
\displaystyle b^i(\vu)  = \frac{\partial_i Q(\vu)}{\partial_i
B(\vu)}= \frac{\partial_i S(\vu)}{\partial_i N(\vu)},\quad\quad
n^i(\vu)= \frac{\partial_i P(\vu)}{\partial_i B(\vu)}=
\frac{\partial_i R(\vu)}{\partial_i N(\vu)}.
\end{array}
\end{equation}
Inserting (\ref{clofl}) into the right hand side of
(\ref{bnflow}), we easily get
\begin{equation}\label{bnflow1}
n^{i} (\vu) = { \nabla}^i{ \nabla}_i N + \sum_{(l)} \epsilon_{(l)}
Z^{(l)} w^i_{(l)},\quad b^{i} (\vu) = { \nabla}^i{ \nabla}_i B +
\sum_{(l)} \epsilon_{(l)} T^{(l)} w^i_{(l)},
\end{equation}
Moreover, using (\ref{clofor}), it is easy to verify that the
reciprocal auxiliary flows
\[
u^i_{{\hat \tau}} = {\hat n}^i(\vu) u^i_{{\hat x}}, \quad\quad
u^i_{{\hat \zeta}} = {\hat b}^i(\vu) u^i_{{\hat x}}, \quad\quad
u^i_{{\hat t}^{(l)}} = {\hat w}^i_{(l)} (\vu) u^i_{{\hat x}},
\]
satisfy
\begin{equation}\label{travel}
\begin{array}{l}
\displaystyle {\hat n}^i(\vu) = \left( n^i B - P + (N n^i-R){\hat
v}^i \right) = \left( \frac{{ H}_i}{{\hat H}_i}n^i-P-{\hat
v}^iR\right) ,
\\
\displaystyle {\hat b}^i(\vu) = \left( b^i B - Q + (Nb^i-S){\hat
v}^i \right) =\left( \frac{{ H}_i}{{\hat H}_i}b^i-Q-{\hat
v}^iS\right),
\\
\displaystyle {\hat w}^i_{(l)}(\vu) = \left( w^i_{(l)} B - T^{(l)}
+ (Nw^i_{(l)}-Z^{(l)}){\hat v}^i \right)=\left( \frac{{
H}_i}{{\hat H}_i}w^i_{(l)}-T^{(l)}-{\hat v}^iZ^{(l)}\right) .
\end{array}
\end{equation}
Finally,
\begin{equation}\label{grads}
\begin{array}{c}
\displaystyle \left( { \nabla} B \right)^2 = 2Q
-\sum_{l}\epsilon_{(l)}\left( T^{(l)} \right) ^2, \quad\quad\left(
{ \nabla} N \right)^2 = 2R -\sum_{l}\epsilon_{(l)}\left( Z^{(l)}
\right) ^2,
\\
\displaystyle < { \nabla} N, { \nabla} B > =
\sum_{l}\epsilon_{(l)}T^{(l)} Z^{(l)}- P-S.
\end{array}
\end{equation}

Then inserting, (\ref{clofl}-\ref{grads}) into the expression of
${\hat R}^{ij}_{ij}(\vu)$, we get the following

\begin{theorem}\label{theo3.2}
Let $g^{ii} (\vu)$ be the metric for the Hamiltonian hydrodynamic
system (\ref{sisdia}), with Christoffel symbols
$\Gamma^{i}_{jk}(\vu)$ and affinors $w^{i}_{(l)}$. Then, after the
reciprocal transformation (\ref{xt}), the non zero components of
the reciprocal Riemannian curvature tensor are
\[
{\hat R}^{ij}_{ij} (\vu) = \sum_{l}\epsilon^{(l)} {\hat
w}^i_{(l)}(\vu) {\hat w}^j_{(l)}(\vu) + {\hat v}^i(\vu) {\hat
n}^j(\vu) +{\hat v}^j (\vu){\hat n}^i(\vu) + {\hat b}^i (\vu)+
{\hat b}^j(\vu), \quad\quad i\not = j,
\]
and the reciprocal Hamiltonian operator takes the form
\begin{equation}\label{trapoi}
\begin{array}{rcl}
{\hat J}^{ij}(\vu) &=&\displaystyle {\hat g}^{ii}(\vu)
\left(\delta^i_j \frac{d}{d{\hat x}}- {\hat \Gamma}^{j}_{ik}(\vu)
u^k_{\hat x}\right)+\sum_{l}\epsilon^{(l)} {\hat w}^i_{(l)}(\vu)
u^i_{\hat x} \left( \frac{d}{d{\hat x}} \right)^{-1} {\hat
w}^j_{(l)}(\vu)  u^j_{\hat x} \\
&&\displaystyle +{\hat b}^i(\vu) u^i_{\hat x} \left(
\frac{d}{d{\hat x}} \right)^{-1} u^j_{\hat x}+ u^i_{\hat x} \left(
\frac{d}{d{\hat x}} \right)^{-1} {\hat b}^j(\vu)
u^j_{\hat x}\\
&&\displaystyle +{\hat n}^i (\vu)u^i_{\hat x} \left(
\frac{d}{d{\hat x}} \right)^{-1} {\hat v}^j (\vu)u^j_{\hat
x}+{\hat v}^i(\vu) u^i_{\hat x} \left( \frac{d}{d{\hat x}}
\right)^{-1} {\hat n}^j (\vu)u^j_{\hat x},
\end{array}
\end{equation}
where the reciprocal metric ${\hat g}^{ii}(\vu)=1/{\hat g}_{ii}(\vu)$ and the
reciprocal affinors ${\hat n}^i(\vu)$, ${\hat b}^i(\vu)$ and
${\hat w}^i_{(l)}(\vu)$ have been defined in (\ref{tramet}) and
(\ref{travel}), respectively.
\end{theorem}

\begin{corollary}
In the special case, when the reciprocal transformation changes
only $x$ ($N(\vu)=0$ and $M(\vu)=1$ in (\ref{xt})), then the
nonzero components of the transformed curvature tensor take the
form
\begin{equation}\label{curvx}
\begin{array}{rl}
\displaystyle {\hat R}^{ij}_{ij} (\vu) &=\; \displaystyle B^2(\vu)
{ R}^{ij}_{ij} (\vu)+ B (\vu)({ \nabla}^i{ \nabla}_i B(\vu)+ {
\nabla}^j{ \nabla}_j B(\vu)) - ({ \nabla} B(\vu))^2 \\
&\; =\displaystyle \sum_{l}\epsilon^{(l)} {\hat w}^i_{(l)}(\vu)
{\hat w}^j_{(l)}(\vu) + {\hat b}^i (\vu)+ {\hat b}^j(\vu).
\end{array}
\end{equation}
In the special case, when the reciprocal transformation changes
only $t$ ($B(\vu)=1$ and $A(\vu)=0$ in (\ref{xt})), then the
nonzero components of the transformed curvature tensor satisfy
\begin{equation}\label{curvt}
\begin{array}{rl}
\displaystyle {\hat R}^{ij}_{ij} (\vu) &= \;\displaystyle\frac{M^2
{ R}^{ij}_{ij} + M \, (v^j\, { \nabla}^i{ \nabla}_i N+ v^i\, {
\nabla}^j{ \nabla}_j N) - v^i\, v^j\, ({ \nabla} N)^2}{(M-Nv^i)
(M-Nv^j)}
\\
&\displaystyle = \;\sum_{l}\epsilon^{(l)} {\hat
w}^i_{(l)}(\vu){\hat w}^j_{(l)}(\vu)+ {\hat v}^i(\vu) {\hat
n}^j(\vu) +{\hat v}^j(\vu) {\hat n}^i(\vu).
\end{array}
\end{equation}
\end{corollary}

\section{On bi--hamiltonicity of the reciprocal system}

Bi--hamiltonicity \cite{Magri} is a relevant property for a
Hamiltonian system (see \cite{Du2,Du0,DLZ} and references
therein). In this section, we suppose that the initial
hydrodynamic system (\ref{sisdia}) $u^i_t= v^i(\vu)u^i_x$,
$i=1,\dots, n$, possesses a bi--hamiltonian structure, that is, it
possesses two non--degenerate compatible Poisson structure
$J^{ij}_{\alpha} (\vu)$, $\alpha=1,2$ and prove that the
reciprocal system is still bi--hamiltonian.

We recall that the Poisson structures $J^{ij}_{1} (\vu)$ and
$J^{ij}_{2} (\vu)$ are compatible if the linear combination
\[
J^{ij}_{1} (\vu)+\lb J^{ij}_{2} (\vu)
\]
is a non degenerate Poisson structure for arbitrary  constant
$\lb$. Let us suppose that the diagonal metrics
$g^{ii}_{(\alpha)}$ are associated to the Poisson structures
$J^{ij}_{\alpha} (\vu)$, $\alpha=1,2$ of the form (\ref{DN}) or
(\ref{MF}). If the second metric $g^{ii}_{(2)}$ is of the form
$g^{ii}_{(2)}=g^{ii}_{(1)}f^i(u^i)$, where $f^i(u^i)$ is an
arbitrary function of one variable, then $J^{ij}_{1} (\vu)+\lb
J^{ij}_{2} (\vu)$ is a Poisson operator associated to the metric
\cite{Mok01}
\[
g^{ii}_{(1)} (\vu) +\lb g^{ii}_{(2)} (\vu),
\]
for arbitrary constants $\lb$.

A natural question is whether, reciprocal transformations preserve
the compatibility of Poisson brackets. Under the action of a
linear reciprocal transformation, ${\hat t} = b x + a t$, ${\hat
x} = n x + m t$, with $a,b,m,n$ constants such that
$(b\,m-a\,n)\not = 0$, the reciprocal to a local Hamiltonian
structure is local (see~\cite{Tsa,Pav}) and bi--hamiltonicity is
preserved~\cite{Mok01},\cite{XZ}.

\begin{theorem}
Suppose that the hydrodynamic system (\ref{sisdia}), $u^i_t=
v^i(\vu)u^i_x$, $i=1,\dots, n$, possesses a bi--hamiltonian
structure such that the associated metrics are non--singular,
diagonal and of the form $g^{ii}_{(1)}$ and
$g^{ii}_{(2)}=g^{ii}_{(1)}f^i(u^i)$.  Then, after the
transformation (\ref{tramet}), the reciprocal system $u^i_{{\hat
t}}= {\hat v}^i(\vu)u^i_{{\hat x}}$, $i=1,\dots, n$, still
possesses a (possibly non--local) bi--hamiltonian structure.
\end{theorem}
To prove the theorem it is sufficient to show that the
corresponding transformed Poisson operators $\hat{J}^{ij}_{1}
(\vu)$ and $\hat{J}^{ij}_{2} (\vu)$  of the form (\ref{trapoi})
are compatible namely
\[
\hat{J}^{ij}_{\lb}=\hat{J}^{ij}_{1} (\vu)+\lb\hat{J}^{ij}_{2} (\vu),
\]
is an Hamiltonian operator associated to the metric
\begin{equation}
\label{med}
\hat{g}^{ii}_{\lb}=\hat{g}^{ii}_{(1)}+\lb\hat{g}^{ii}_{(1)}f^i(u^i).
\end{equation}
It is straightforward to show that the local part of the operator
$\hat{J}^{ij}_{\lb}$ is linear in $\lb$. In order to show that the
nonlocal part is also linear is $\lb$ it is sufficient to use a
result of \cite{Mok01} which says that the curvature tensor of a
metric of the form $\hat{g}^{ii}_{\lb}$ defined in (\ref{med}) is
linear in $\lb$.

In the next sections we consider hydrodynamic type system
(\ref{sisdia}) with a Hamiltonian structure associated to either a
flat or a constant curvature or a conformally flat metric
$g^{ii}_{(\alpha)}(\vu)$, $\alpha=1,2$, and we give conditions for
the flatness of the reciprocal metric ${\hat g}^{ii}_{(\alpha)}
(\vu)$. If ${\hat g}^{ii}_{(\alpha)} (\vu)$, $\alpha=1,2$ are both
flat and the initial system is bi--hamiltonian, then, by the
theorem above, the reciprocal system possesses a flat
bi--hamiltonian structure.

\section{Conditions for reciprocal flat metrics when only $x$ changes}\label{sec5}

In this and the following sections, we suppose that the initial
hydrodynamic system $u^i_t = v^i(\vu) u^i_x$, $i=1,\dots, n$ is
Hamiltonian as in (\ref{sisdia}) and the associated Hamiltonian
operator $J^{ij}(\vu)$ is as in (\ref{DN}) or in (\ref{MF}), and
we look for necessary and sufficient conditions such that, after a
reciprocal transformation of type (\ref{xt}), one of the
reciprocal metrics be flat. Since the reciprocal transformation
$d{\hat x} = B (\vu)dx + A(\vu)dt$, $d{\hat t} = N(\vu)dx
+M(\vu)dt,$ is the composition of a transformation of the variable
$x$ and of a transformation of the variable $t$, we start our
investigation with reciprocal transformations in which only the
$x$ variable changes.

In this section, using notations settled in sections 2 and 3, we
give the complete set of necessary and sufficient conditions for
the reciprocal metric to be flat, when $n\ge 3$ and the initial
non--singular metric $g^{ii} (\vu)$ is either flat or of constant
curvature $c$ or conformally flat with affinors $w^i(\vu)$. Then
the extended reciprocal transformation (\ref{clofor}) is
\begin{equation}\label{extx}
d{\hat x} = B (\vu)dx + A(\vu)dt+ P(\vu)d\tau + Q(\vu) d\zeta + T
(\vu)dt_{w}, \; {\hat t} = t, \; {\hat \tau} =\tau; \; {\hat t}_w
= t_w.
\end{equation}
\begin{remark}\label{rem1}
Let $b^i(\vu)$ and  $w^i(\vu)$ be  as in (\ref{bnflow}).
The quantities $Q(\vu)$ and $T(\vu)$ in (\ref{extx})
satisfy the relations
\[
b^i (\vu)=\displaystyle \frac{\partial_i Q(\vu)}{\partial_i B(\vu)},\quad
w^i
(\vu) =\displaystyle \frac{\partial_i T(\vu)}{\partial_i B(\vu)}.
\]
Since  (\ref{extx}) is a closed form, $Q(\vu)$ and $T(\vu)$ satisfy the relation
\begin{equation}\label{Qx}
Q(\vu) =\displaystyle \frac{1}{2} \Big(\nabla B (\vu)\Big)^2
+B(\vu) T(\vu).
\end{equation}
Note that
\begin{itemize}
\item $T(\vu)=0$ if  $g^{ii}(\vu)$ is flat;
\item $T(\vu)=\frac{c}{2} B(\vu)$ if $g^{ii}(\vu)$ is of constant curvature $c$.
\end{itemize}
\end{remark}

After the reciprocal transformation (\ref{extx}), the metric ${\hat g}^{ii}
(\vu)$ reciprocal to $g^{ii} (\vu)$, is flat if and only if the r.h.s.
in (\ref{curvx}) is zero,
$\forall i,k=1,\dots,n$, $i\not =k$, that is
\begin{equation}\label{recx0}
\displaystyle B^2(\vu) (w^i(\vu) +w^k(\vu))+ B (\vu)({ \nabla}^i{
\nabla}_i B(\vu)+ { \nabla}^k{ \nabla}_k B(\vu)) - ({ \nabla}
B(\vu))^2 \equiv {\hat b}^i (\vu) +{\hat b}^k (\vu) = 0.
\end{equation}
(\ref{recx0}) depends only on the initial Poisson
structure and on the density of conservation law $B(\vu)$ in the
reciprocal transformation. The above formula also shows that the
class of metrics which are either flat or of constant curvature or
conformally flat is left invariant by reciprocal transformations
of the independent variable $x$ and that ${\hat g}^{ii} (\vu)$ is
flat if and only if the reciprocal affinor ${\hat b}^i (\vu)\equiv
0$, $i=1,\dots, n$. The next theorem gives the necessary and
sufficient conditions for ${\hat b}^{i}(\vu)\equiv 0$ in function of the
initial system.

\begin{theorem}\label{equiv}
Let the contravariant non--singular diagonal  metric  $g^{ii}
(\vu)$ associated to the initial system (\ref{sisdia}) be either
flat or of constant curvature $c$ or conformally flat with
affinors $w^i(\vu)$. Let $d{\hat x} = B (\vu)dx + A(\vu)dt$,
$d{\hat t} = dt$, with $B(\vu)\not \equiv const.$. Then the
reciprocal metric ${\hat g}^{ii} (\vu)$ is flat if and only if
there exists a constant $\kappa$ such that
\begin{equation}\label{nablab}
\frac{Q(\vu)}{B(\vu)} \equiv \frac{\left( \nabla
B(\vu)\right)^2}{2B(\vu)} + T(\vu)= \kappa,
\end{equation}
where $Q(\vu)$ and $T(\vu)$ are as in Remark~\ref{rem1}.
If in (\ref{nablab})  $\kappa=0$, then $B(\vu)$ is a Casimir
associated to the metric $g^{ii} (\vu)$; if $\kappa\not =0$ in
(\ref{nablab}) then $B(\vu)$ is proportional to a  density of
momentum associated to the metric $g^{ii} (\vu)$.
\end{theorem}

{\it Proof.} \hspace{ 0.5 truecm} Since
\begin{equation}\label{flatflow}
\partial_i
\left( \nabla B (\vu) \right)^2 = 2\partial_i B (\vu) \,\nabla^i
\nabla_i B(\vu),
\end{equation}
(\ref{recx0}) is equivalent to
\begin{equation}\label{nablab1}
\begin{array}{ll}
0 &=\displaystyle B^2 (\vu) w^i(\vu) +B(\vu)\nabla^i\nabla_i
B(\vu) -\frac{1}{2} \left( \nabla B(\vu)\right)^2 = B (\vu)
b^i(\vu) -Q(\vu) \\
&\displaystyle = \frac{B^2(\vu)}{\partial_i B(\vu)} \partial_i
\left( \frac{Q(\vu)}{B(\vu)}\right), \quad\quad i=1,\dots,n,
\end{array}
\end{equation}
where $Q(\vu)$ is as in (\ref{Qx}) and statement (\ref{nablab})
immediately follows.

Finally, inserting (\ref{nablab}) into the expression of the
auxiliary flow $b^i(\vu)$, we get
\[
b^i(\vu) \equiv \frac{\partial_i Q(\vu)}{\partial_i B(\vu)} =
\kappa, \quad\quad i=1,\dots, n.\quad\quad \square
\]

(\ref{nablab}) settles quite restrictive conditions on
the density of conservation law $B(\vu)$ in the reciprocal
transformation for which we may hope that the reciprocal metric be flat.
The following theorem shows that, conversely, if
$B(\vu)$ is either a Casimir or a density of momentum associated
to the metric $g^{ii} (\vu)$ then the reciprocal metric ${\hat
g}^{ii} (\vu)$ is, at worse, of constant curvature.

\begin{theorem}
Let the contravariant non--singular diagonal  metric  $g^{ii}
(\vu)$ associated to the initial system (\ref{sisdia}) be either
flat or of constant curvature $c$ or conformally flat with
affinors $w^i(\vu)$ and $T(\vu)$ as in Remark \ref{rem1}. Let
$B(\vu)$ be either a Casimir ($b=0$) or a density of momentum
($b=1$) associated to the metric $g^{ii}(\vu)$. Then, under the
reciprocal transformation $d{\hat x} = B (\vu)dx + A(\vu)dt$,
$d{\hat t} = dt$, the reciprocal metric ${\hat g}^{ii} (\vu)$ is
either flat or of constant curvature ${\hat c}$ where
\begin{equation}\label{curvc}
{\hat c} = 2b B(\vu) -2B(\vu) T(\vu) - \left( \nabla B(\vu)
\right)^2.
\end{equation}
\end{theorem}

{\it Proof.} \hspace{.5 truecm} Let $B(\vu)$ be as in the
hypothesis, then (\ref{curvx}) becomes
\[
\begin{array}{ll}
\displaystyle{\hat R}^{ik}_{ik} (\vu) &= \displaystyle B^2(\vu)
\Big(w^i(\vu) +w^k(\vu) \Big) + B(\vu) \Big(\nabla^i\nabla_i
B(\vu) + \nabla^k\nabla_k B(\vu) \Big) - \Big( \nabla
B(\vu)\Big)^2 \\
&\\
&\displaystyle = 2b B(\vu)-2 B(\vu) T(\vu) -\Big( \nabla
B(\vu)\Big)^2,
\end{array}
\]
and, for $l=1,\dots, n$, we have
\[
\partial_l \left( 2b B(\vu)-2 B(\vu) T(\vu)
-\Big( \nabla B(\vu)\Big)^2 \right) = 2 B(\vu) \Big(
w^l(\vu)\partial_l B(\vu) - \partial_l T(\vu) \Big) =0,
\]
from which we conclude that ${\hat R}^{ik}_{ik} (\vu)$ is a
constant function. $\quad\quad \square$

The above necessary and sufficient conditions take a particular
simple form in the case in which the initial metric is flat:

\begin{corollary}
Let $g^{ii} (\vu)$ be the flat metric for DN system (\ref{DN}) and
$d{\hat x} =B(\vu)dx+A(\vu)dt$ be a reciprocal transformation.
Then the reciprocal metric ${\hat g}^{ii} (\vu) = B^2 (\vu) g^{ii}
(\vu)$ is flat if and only if one of the following conditions hold
true:

1) $B$ and $A$ are constant functions;

2) $B(\vu)$ is a Casimir of the metric $g^{ii} (\vu)$ and
$(\nabla B(\vu))^2=0$;

3) $B(\vu)$ is a density of momentum for the metric $g^{ii} (\vu)$ and
$(\nabla B(\vu))^2=2B(\vu)$.
\end{corollary}

\begin{remark}
If $B(\vu)$ and $N(\vu)$ are non trivial independent Casimirs of
the flat metric $g^{ii} (\vu)$ and $(\nabla B (\vu))^2\not =0$,
then there exist a constant $\alpha$ and
$A(\vu)$ such that, under the reciprocal transformation $d{\hat x}
= (\alpha B(\vu)+N(\vu)) dx + A(\vu) dt$, the reciprocal metric ${\hat
g}^{ii} (\vu) = (\alpha B(\vu)+N(\vu))^2 g^{ii} (\vu)$ is flat.

If $B(\vu)$ is a density of momentum for the flat metric
$g^{ii} (\vu)$ and $(\nabla B(\vu))^2-2B(\vu)=2\alpha$, then under
the reciprocal transformation $d{\hat x} = (B(\vu)+\alpha) dx +
A(\vu) dt$, the reciprocal metric ${\hat g}^{ii} (\vu) =
(B(\vu)+\alpha)^2 g^{ii} (\vu)$ is flat.
\end{remark}

\section{Conditions for reciprocal flat metrics when only $t$ changes}

In this section, we give the complete set of necessary and
sufficient conditions for a flat reciprocal metric, when $n\ge 3$,
$g^{ii} (\vu)$ is the flat metric associated to the initial DN
hydrodynamic type system
\begin{equation}\label{sss}
u^i_t = v^i(\vu) u^i_x=J^{ij}(\vu) \partial_i H(\vu),
\end{equation}
and the reciprocal
transformation is ${\hat x} = x$, $d{\hat t} = N(\vu) dx + M(\vu)
dt$. Under these hypotheses, the reciprocal metric ${\hat g}^{ii}
(\vu)$ is flat if and only if the r.h.s. in (\ref{curvt}) is identically zero
$ \forall i,k, \,\; i\not =k$, that is
\begin{equation}\label{rect0}
\displaystyle  M (\vu)\, \Big(v^j(\vu)\, { \nabla}^i{ \nabla}_i
N(\vu)+ v^i(\vu)\, { \nabla}^j{ \nabla}_j N(\vu)\Big) - v^i(\vu)\,
v^j(\vu)\, ({ \nabla} N)^2(\vu)\equiv 0.
\end{equation}
(\ref{rect0}) explicitly depends on the initial Poisson structure,
on the density of conservation law $N(\vu)$ in the reciprocal
transformation and on the density of Hamiltonian $H(\vu)$
associated to the metric $g^{ii}(\vu)$.

\begin{theorem}\label{equivt}
Let $g^{ii} (\vu)$ be the diagonal non--degenerate flat metric for
(\ref{DN}) and let $v^i(\vu)\not = const.$, $i=1,\dots,n$. Let
$d{\hat t} = N (\vu)dx + M(\vu)dt$, $d{\hat x} = dx$, with
$N(\vu)\not \equiv const.$. Then the reciprocal metric ${\hat
g}^{ii} (\vu)$ is flat if and only if there exists a constant
$\kappa$ such that
\begin{equation}\label{nablan}
\frac{\left( \nabla N(\vu)\right)^2}{2M(\vu)} = \kappa.
\end{equation}
If $\kappa=0$, then $N(\vu)$ is a Casimir
associated to the metric $g^{ii} (\vu)$; if $\kappa\not =0$,
 then $N(\vu)=\kappa H(\vu)$, where $H(\vu)$ is a
density of Hamiltonian associated to the metric $g^{ii} (\vu)$.
\end{theorem}

{\it Proof.} \hspace{ 0.5 truecm} (\ref{rect0}) is equivalent to
\[
\begin{array}{ll}
0 &=\displaystyle M (\vu) \left( v^k(\vu)n^i(\vu)
+v^i(\vu)n^k(\vu) \right) - v^i(\vu) v^k(\vu)\left( \nabla
N(\vu)\right)^2 \\
&\displaystyle = \frac{M^2(\vu)}{\partial_i N(\vu)\partial_k
N(\vu)}\Bigg( \partial_i M(\vu) \partial_k \left( \frac{\Big(
\nabla N(\vu)\Big)^2}{2M(\vu)}\right) +\partial_k M(\vu)
\partial_i \left( \frac{\Big( \nabla N(\vu)\Big)^2}{2M(\vu)}\right)\Bigg),
\end{array}
\]
$\forall i,k=1,\dots,n,\; i\not=k$. If (\ref{nablan}) holds, then
${\hat R}^{ik}_{ik} \equiv 0$, $i,k=1,\dots,n$, $i\not =k$, and
$n^i(\vu)$ is either the null velocity flow ($N(\vu)$ Casimir of
the initial metric) or
\[
n^i(\vu) \equiv \frac{\partial_i \Big( \nabla
N(\vu)\Big)^2}{2\partial_i N(\vu)} = \kappa\frac{\partial_i
M(\vu)}{\partial_i N(\vu)} = \kappa v^i(\vu), \quad\quad
i=1,\dots, n.
\]
Viceversa, suppose that the reciprocal metric is flat and $\Big(
\nabla N(\vu)\Big)^2\not=0$, then $\forall i,k,l=1,\dots,n$,
$i\not=k$, it is straightforward to verify
\[
\begin{array}{ll}
0&\displaystyle\equiv \partial_l \left( M(\vu) \Big( v^i(\vu)
n^k(\vu) + v^k(\vu) n^i(\vu)\Big) -v^i(\vu) v^k(\vu)\Big( \nabla
N(\vu)\Big)^2\right) \\
&\displaystyle = \partial_l M(\vu) \Big( v^i (\vu) n^k(\vu) +
v^k(\vu) n^i (\vu) \Big) -v^i(\vu) v^k(\vu) \Big( \nabla
N(\vu)\Big)^2 \\
&\displaystyle =-v^i(\vu) v^k(\vu) \Big( \nabla N(\vu)\Big)^2
\partial_l \log \left( \frac{\Big( \nabla
N(\vu)\Big)^2}{M(\vu)}\right).\quad \square
\end{array}
\]

Equation (\ref{nablan}) settles quite restrictive conditions on
the density of conservation law $N(\vu)$ in order to
preserve flatness of the metric.
The following theorem shows that, conversely, if
$N(\vu)$ is either a Casimir or a density of Hamiltonian
associated to the flat metric $g^{ii} (\vu)$ then the reciprocal
metric ${\hat g}^{ii} (\vu)$ is either flat or associated to an
hypersurface in the Euclidean space.

\begin{theorem}
Let $g^{ii} (\vu)$ be a contravariant flat non--singular diagonal
metric for (\ref{sss}) and let $N(\vu)$ be either a Casimir or a
density of Hamiltonian associated to the metric $g^{ii}(\vu)$.
Then, under the reciprocal transformation $d{\hat t} = N (\vu)dx +
M(\vu)dt$, $d{\hat x} = dx$, the reciprocal metric ${\hat g}^{ii}
(\vu)$ is either flat or the reciprocal Poisson operator takes the
form
\begin{equation}\label{hyper}
{\hat J}^{ij} (\vu) = {\hat g}^{ii}(\vu) \delta^j_i
\frac{d}{d{\hat x}} - {\hat g}^{ii} (\vu) {\hat \Gamma}^{j}_{ik}
u^k_{{\hat x}} + \gamma {\hat v}^iu^i_{{\hat x}}
\left(\frac{d}{d{\hat x}}\right)^{-1} {\hat v}^j_x u^j_x,
\end{equation}
with $\gamma$ constant.
\end{theorem}

{\it Proof.}  If $N(\vu)$ is a Casimir and $\Big( \nabla
N(\vu)\Big)^2=\gamma$, then ${\hat R}^{ik}_{ik} (\vu) = -\gamma{\hat
v}^i(\vu) {\hat v}^k(\vu)$, $\forall i,k=1,\dots,n$, $i\not =k$
and the assertion easily follows.

If $N(\vu)$ is a density of Hamiltonian and
$M(\vu)=\displaystyle \frac{1}{2}\Big( \nabla N(\vu)\Big)^2 +\gamma$,
then ${\hat R}^{ik}_{ik} (\vu)= \gamma {\hat v}^i(\vu) {\hat
v}^k(\vu)$, $\forall i,k=1,\dots,n$, $i\not =k$, and the assertion
easily follows.$\square$

\begin{remark}
If $B(\vu)$ and $N(\vu)$ are non trivial independent Casimirs of
the flat metric $g^{ii} (\vu)$ and $(\nabla N (\vu))^2\not =0$,
then there exist a constant $\alpha$ and
$M(\vu)$ such that, under the reciprocal transformation $d{\hat t}
= (\alpha N(\vu)+B(\vu)) dx + M(\vu) dt$, the reciprocal metric ${\hat
g}^{ii} (\vu)$ is flat.

If $N(\vu)$ is a density of Hamiltonian for the flat metric
$g^{ii} (\vu)$ and $(\nabla N(\vu))^2-2M(\vu)=2\alpha$, then under
the reciprocal transformation $d{\hat t} = N(\vu) dx +
\Big(M(\vu)+\alpha\big) dt$, the reciprocal metric ${\hat g}^{ii}
(\vu) $ is flat.
\end{remark}

\section{Conditions for flat reciprocal metrics when
the transformation changes both $x$ and $t$}

Let the initial hydrodynamic system be Hamiltonian as in
(\ref{sisdia})
\begin{equation}\label{sss1}
u^i_t = v^i(\vu) u^i_x=J^{ij}(\vu) \partial_i H(\vu),
\end{equation}
where $J^{ij}(\vu)$ the Hamiltonian operator as in (\ref{DN}) or
in (\ref{MF}) is associated to a initial metric $g^{ii}(\vu)$
either flat or of constant curvature $c$ or conformally flat with
affinors $w^i(\vu)$.

The reciprocal transformation
\[
d{\hat x} = B (\vu)dx + A(\vu)dt,\quad\quad d{\hat t} = N(\vu)dx +
M(\vu)dt,
\]
is the composition of the following two reciprocal
transformations of one variable
\[
d{\tilde x} = B (\vu)dx + A(\vu)dt,\quad\quad d{\tilde t} = dt,
\]
\[
d{\hat x} =d{\tilde x},\quad\quad d{\hat t} = {\tilde
N}(\vu)d{\tilde x} + {\tilde M}(\vu)d{\tilde t},
\]
where
\begin{equation}\label{tildeN}
{\tilde N} (\vu) = \frac{N(\vu)}{B(\vu)},\quad\quad {\tilde M}
(\vu) = \frac{M(\vu)B(\vu)-N(\vu)A(\vu)}{B(\vu)}.
\end{equation}
In view of the results of the previous sections, it is natural to
restrict the attention to the case in which $B(\vu)$ is either a
Casimir or a momentum density associated to the metric
$g_{ii}(\vu)$. Then, after the first reciprocal transformation,
the metric ${\tilde g}_{ii}(\vu)$ is either flat (${\hat c} =0$)
or of constant curvature ${\hat c}\not =0$, where ${\hat c}$ is
the expression in the right hand side of (\ref{curvc}). Let ${\hat
c} =0$, then after the second reciprocal transformation, by
Theorem~\ref{equivt} the metric ${\hat g}_{ii}(\vu)$ is flat if
and only if there exists a constant ${\tilde \kappa}$ such that
\begin{equation}\label{nablaxt}
\frac{\left( {\tilde \nabla} {\tilde N} (\vu) \right)^2}{2{\tilde
M}(\vu)}  = {\tilde \kappa}.
\end{equation}
We want to express (\ref{nablaxt}) in an  equivalent way as a  function
of the initial metric $g^{ii} (\vu)$ and of the density of
conservation laws $N(\vu)$ and $B(\vu)$.

\begin{remark}\label{rem2}
Let $n^i(\vu)$ and  $w^i(\vu)$ be  as in (\ref{bnflow}).
The quantities $R(\vu)$ and $Z(\vu)$ in (\ref{clofor})
satisfy the relations
\[
n^i (\vu)=\displaystyle \frac{\partial_i R(\vu)}{\partial_i N(\vu)},\quad
w^i
(\vu) =\displaystyle \frac{\partial_i Z(\vu)}{\partial_i N(\vu)}.
\]
Since  (\ref{clofor}) is a closed form, $R(\vu)$ and $Z(\vu)$ satisfy the relation
\begin{equation}\label{rconv}
R(\vu)
=\displaystyle \frac{1}{2} \Big(\nabla N (\vu)\Big)^2 +N(\vu)
Z(\vu).
\end{equation}
Note that
\begin{itemize}
\item $Z(\vu)=0$ if  $g^{ii}(\vu)$ is flat;
\item $Z(\vu)=\frac{c}{2} N(\vu)$ if $g^{ii}(\vu)$ is of constant curvature $c$.
\end{itemize}
\end{remark}

Inserting (\ref{tildeN}) into (\ref{nablaxt}), we get
\begin{equation}\label{n1}
\begin{array}{l}
\displaystyle {\tilde \kappa} \left(M(\vu) -
\frac{N(\vu)}{B(\vu)}A(\vu)\right) -\frac{1}{2}\Big(\nabla
N(\vu)\Big)^2 - \frac{N^2(\vu)}{2B^2(\vu)}\Big(\nabla
B(\vu)\Big)^2\\
\displaystyle\quad\quad +\frac{N(\vu)}{B(\vu)}<\nabla
B(\vu),\nabla N(\vu)> =0.
\end{array}
\end{equation}
If ${\tilde \kappa} =0$ in (\ref{n1}), then either $N(\vu)= \nu_1
B(\vu)$, $M(\vu) = \nu_1 A(\vu) +\nu_2$, with $\nu_1,\nu_2$
non--zero constants, or there exists a constant $\nu_3$ such that
\begin{equation}\label{n2}
\frac{(\nabla N(\vu))^2}{2N(\vu)} +Z(\vu) = \nu_3.
\end{equation}
Comparing (\ref{n2}) with (\ref{nablan}) and (\ref{rconv}), we
conclude that $N(\vu)$ is either a Casimir ($\nu_3=0$) or
proportional to a momentum density $(\nu_3\not =0)$ for the
initial metric $g^{ii}(\vu)$.

If ${\tilde \kappa} \not =0$ in (\ref{n1}), then there exists of a
constant $\nu_4$ such that
\[
\frac{(\nabla N(\vu))^2}{2} +Z(\vu)N(\vu) = {\tilde \kappa} M(\vu)
+\nu_4 N(\vu),
\]
that is $N(\vu)$ is the linear combination with constant
coefficients of a Hamiltonian density $H(\vu)$ and a momentum
density associated to the initial metric $g^{ii} (\vu)$.

In the next theorem we summarize the above discussion. We use the
notations settled in remarks \ref{rem1} and \ref{rem2}.

\begin{theorem}\label{theo7.2}
Let the non--singular metric $g^{ii}(\vu)$ of system (\ref{sss1})
be either flat or of constant curvature $c$ or
conformally flat  with affinors $w^i(\vu)$.
Let $B(\vu)$ be either a Casimir ($b=0$) or a momentum density
($b=1$) associated to the metric $g_{ii}(\vu)$ and such that
\begin{equation}\label{fflat2}
\displaystyle \frac{\left( \nabla B(\vu)\right)^2}{2B(\vu)} +
T(\vu)= b,
\end{equation}
where $T(\vu)$ has been defined in Remark~\ref{rem1}.
Then, after the reciprocal transformation $ d{\hat x} = B (\vu)dx
+ A(\vu)dt$, $d{\hat t} = N(\vu)dx + M(\vu)dt$, the reciprocal
metric ${\hat g}^{ii} (\vu)$ is flat if and only if either there
exist non--zero constants $\nu_1$ and $\nu_2$ such that $N(\vu)=
\nu_1 B(\vu)$, $M(\vu) = \nu_1 A(\vu) +\nu_2$ or there exist
(possibly zero) constants $\nu_3,\nu_4$ such that
\begin{equation}\label{fflat1}
\frac{(\nabla N(\vu))^2}{2} +Z(\vu)N(\vu) = \nu_3 M(\vu) +\nu_4
N(\vu),
\end{equation}
where $Z(\vu)$ has been defined in Remark~\ref{rem2}.
If (\ref{fflat1}) holds true then $N(\vu)$ is either a Casimir
$(\nu_3=\nu_4=0)$ or a momentum density $(\nu_3=0,\nu_4=1)$ or a
Hamiltonian density $(\nu_3=1,\nu_4=0)$ or a linear combination
with constants coefficients of the Casimirs, momentum and
Hamiltonian density for the initial metric $g^{ii}(\vu)$.
\end{theorem}

The above theorem is far from setting the whole set of necessary
and sufficient conditions for the reciprocal metric to be flat. In
next theorem we give another set of sufficient conditions for a
flat reciprocal metric when $B(\vu)=H(\vu)$ is the Hamiltonian
density in (\ref{sss1}). Again we use the notations settled in
Remarks \ref{rem1} and \ref{rem2}.

\begin{theorem}
Let the non--singular metric $g^{ii}(\vu)$ of system (\ref{sss1})
be either flat or of constant curvature $c$ or conformally flat
with affinors $w^i(\vu)$. Let $J^{ij} (\vu)$ be the Hamiltonian
operator associated to $g^{ii}(\vu)$ and let
\[
u^i_t = v^i(\vu) u^i_x = J^{ij} (\vu) \partial_j B(\vu).
\]
Under the reciprocal transformation $ d{\hat x} = B (\vu)dx +
A(\vu)dt$, $d{\hat t} = N(\vu)dx + M(\vu)dt$, the reciprocal
metric ${\hat g}^{ii} (\vu)$ is flat if there exists a constant
$\nu_5$ such that
\begin{equation}\label{flat2}
A(\vu) = \frac{1}{2} (\nabla B(\vu))^2+ T(\vu) B(\vu), \quad
\frac{(\nabla N(\vu))^2}{2N(\vu)} +Z(\vu) = \nu_5,
\end{equation}
where $T(\vu)$ and  $Z(\vu)$ have been defined in
Remark~\ref{rem1} and  Remark~\ref{rem2} respectively. If
(\ref{flat2}) holds true then $N(\vu)$ is either a Casimir
$(\nu_5=0)$ or a momentum density $(\nu_5=1)$ for the initial
metric $g_{ii}(\vu)$.
\end{theorem}

{\it Proof.} Under the hypotheses of the theorem, after the first
reciprocal transformation
\[ d{\tilde x} = B (\vu)dx + A(\vu)dt,\quad\quad d{\tilde t} = dt,
\]
the transformed metric ${\tilde g}^{ii} (\vu)= \displaystyle
g^{ii} (\vu)B^2 (\vu)$ is conformally flat with curvature
tensor \[{\tilde R}^{ik}_{ik} (\vu) = {\tilde b}^i (\vu) + {\tilde
b}^k (\vu).
\]
After the second reciprocal transformation
\[
d{\hat x} =d{\tilde x},\quad\quad d{\hat t} = {\tilde
N}(\vu)d{\tilde x} + {\tilde M}(\vu)d{\tilde t},
\]
with ${\tilde N} (\vu)$ and  ${\tilde M} (\vu)$ as in
(\ref{tildeN}), the transformed metric ${\hat g}^{ii} (\vu)$ has
Riemannian curvature tensor
\[
{\hat R}^{ik}_{ik} (\vu) = \frac{{\tilde M}^2  ({\tilde b}^i +
{\tilde b}^k ) + {\tilde M}  \big( {\tilde b}^i  {\tilde \nabla}^k
{\tilde \nabla}_k {\tilde N}  + {\tilde b}^k  {\tilde \nabla}^k
{\tilde \nabla}_k {\tilde N} \big) -{\tilde b}^i {\tilde b}^k
\big( {\tilde \nabla} {\tilde N}  \big)^2 }{ \big( {\tilde M}  -
{\tilde b}^k {\tilde N}  \big) \big( {\tilde M} - {\tilde b}^i
{\tilde N}  \big)},\] where we have dropped the $\vu$ in the r.h.s.. The condition
${\hat R}^{ik}_{ik} (\vu)\equiv 0$ is satisfied in the above relation if
\begin{equation}\label{trans1}
{\tilde M} (\vu) \partial_k {\tilde N} (\vu) +\frac{1}{2}
\partial_k \Big( {\tilde \nabla} {\tilde N} (\vu) \Big)^2
-\frac{\Big( {\tilde \nabla} {\tilde N} (\vu) \Big)^2}{2 {\tilde
M} (\vu)}\partial_k {\tilde M} (\vu) =0.
\end{equation}
Inserting (\ref{tildeN}) and $\displaystyle {\tilde g}^{ii} (\vu)
= g^{ii} (\vu) B^2(\vu)$ into (\ref{trans1}), we get
\begin{equation}\label{trans2}
\begin{array}{l}
\displaystyle M-\frac{AN}{B}+ B \nabla^k \nabla_k N - N \nabla^k
\nabla_k B - < \nabla B, \nabla N> + \frac{N}{B} \Big( \nabla B
\Big)^2 \\
\displaystyle \quad\quad- \frac{B}{2} \left(\Big( \nabla N \Big)^2
+ \frac{N^2}{B^2} \Big( \nabla B \Big)^2 -2\frac{N}{B} < \nabla B,
\nabla N > \right) \frac{B b^k -A}{MB-NA} \equiv 0.
\end{array}
\end{equation}
Let $A(\vu)$ and $N(\vu)$  be as in (\ref{flat2}), then
\[
M(\vu) = <\nabla B(\vu), \nabla N(\vu) > + N(\vu) T(\vu) + B(\vu)
Z(\vu) -\nu_5 B(\vu),
\]
and (\ref{trans2}) is identically satisfied. $\quad \square$

\begin{remark}
If the initial metric of system (\ref{sss1}) is flat,
(\ref{flat2}) is equivalent to
\[
A (\vu)=\frac{1}{2}\Big( \nabla B(\vu) \Big)^2,
\]
and $N(\vu)$ is either a Casimir such that $\big(\nabla N (\vu)
\big)^2 =0$ or $N(\vu)$ is a momentum density such that
$\big(\nabla N (\vu) \big)^2 - 2\nu_5 N(\vu) =0$.
\end{remark}

\section{Examples: flat metrics on moduli space of hyperelliptic curves}
\label{Hurwitz} Flat metrics on Hurwitz spaces have been studied
by Dubrovin in the framework of Frobenius manifolds \cite{Du1}.
The metrics considered in \cite{Du1} are of Egorov type (see
\cite{Du2,Du1,Krich} and references therein for the role of the
algebro--geometric approach in the theory of hydrodynamic
systems). In this section we restrict ourselves to the moduli
space of hyperelliptic Riemann surfaces and on this space we
derive flat metrics which are not of Egorov type.

Let us consider the  hyperelliptic curves of genus $g$
\begin{equation}
\label{hypercurve}
\mathcal{C}:=\{(z,w)\in\mathbb{C}^2\;|w^2=\prod_{k=1}^{2g+1}(\lb-u^k)\},\quad
u^k\neq u^j,k\neq j.
\end{equation}
The distinct parameters $u^1,\dots,u^{2g+1}$ are the local
coordinates on the moduli space of hyperelliptic curves. On the
Riemann surface $\C$ we define  the meromorphic bidifferential
$W(P,Q)$ as
\begin{equation}\label{W-def}
W(P,Q) := d_P d_Q\log E(P,Q)
\end{equation}
where $E(P,Q)$ is the prime form~\cite{Fay}. $W(P,Q)$ is a
symmetric bi-differential on $\C\times \C$ with  second order pole
at the diagonal $P=Q$ with biresidue $1$ and the properties:
\begin{equation}
\oint_{\alpha_k} W(P,Q) = 0 \;; \quad \oint_{\beta_k} W(P,Q) = 2 \pi i \,
\omega_k(P) \;; \quad k=1,\dots, g \;. \label{W-periods}
\end{equation}
Here $\{\alpha_k,\beta_k\}_{k=1}^g$ is the canonical basis of
cycles on $\C$ and $ \{ \omega_k(P) \}_{k=1}^g$ is the
corresponding set of holomorphic differentials normalized by
$\oint_{\alpha_l}\omega_k=\delta_{kl}, \,k,l=1,\dots,g$. The
dependence of the bidifferential $W$ on branch points of the
Riemann surface is given by the Rauch variational formulas
\cite{KK0}:

\begin{equation}
\frac{d W(P,Q)}{du^j}=\frac{1}{2}W(P,P_j)W(Q,P_j)\;,
\label{W-variation}
\end{equation}
where $W(P,P_j)$ denotes the evaluation of the bidifferential
$W(P,Q)$ at $Q=P_j$ with respect to the standard local parameter
$x_j(Q)=\sqrt{\lb(Q)-u^j}$ near the ramification point $P_j\;:$
\begin{equation}
W(P,P_j) := \left.\frac{W(P,Q)}{dx_j(Q)} \right|_{Q=P_j} \;.
\label{notation}
\end{equation}
We consider the Abelian differentials
\begin{equation}
\label{dp}
dp_s(Q)=-\displaystyle\frac{1}{2s-1}\; \res[P=\infty]
\;\lb(P)^\frac{2s-1}{2}W(Q,P),\quad s=1,2,\dots,
\end{equation}
which are normalized differentials of the second kind with a pole
at infinity of order $2s$ and behaviour
\[
dp_s(Q)=-\displaystyle\frac{dz}{z^{2s}}+\text{regular terms},\quad
Q\ra\infty
\]
where $z=1/\sqrt{\lb}$ is the local coordinate in the neighbourhood of infinity.

\begin{theorem}\cite{Du1}
The diagonal  metrics
\begin{equation}
\label{metric}
g_{ii}^0=\res[Q=P_i]\left[\frac{dp_1^2(P)}{d\lambda}\right](d
u^i)^2 =\displaystyle\frac{1}{2}(dp_1(u^i))^2(du^i)^2
,\quad\;\;g_{ii}^1=\displaystyle\frac{g_{ii}^1}{u^i},
\end{equation}
where $dp_1$ is the differential (\ref{dp}), are compatible flat
metrics on the moduli space of hyperelliptic Riemann surfaces.
\end{theorem}
The correspondent flat coordinates of the metric $g_{ii}^0$ are
the following \cite{Du1}
\begin{equation}
\label{flatc1} h_{0} = -\res[\infty] \lambda^{\frac{1}{2}}
dp_1,\;\; r^a = \displaystyle\frac{1}{2\pi
i}\oint_{\beta_a}dp_1,\;\; s_0^a=-\frac{1}{ 2\pi
i}\oint_{\alpha_a}\lambda dp_1,~ a=1,\dots ,g.
\end{equation}
The flat coordinates of the metric $g_{kk}^1$ are obtained by the
relations
\begin{equation}
\label{flatc2} p_1(0),\;\; r^a = \displaystyle\frac{1}{2\pi
i}\oint_{\beta_a}dp_1,\;\;s_1^a=-\frac{1}{ 2\pi
i}\oint_{\alpha_a}\log\lambda dp_1,~ a=1,\dots ,g.
\end{equation}
We observe that the coordinates $  r^a,\,a=1,\dots, g,$
are the common Casimirs of the  metrics $g_{ii}^0$ and $g_{ii}^1$.

Let  $J^{ij}_0$ and $J^{ij}_1$ be the Hamiltonian operators
associated to the metrics $(g^0)^{ii}$ and $(g^1)^{ii}$
respectively and let us consider the Hamiltonian densities
\begin{equation}
\label{kdvh} h_s = -\res[\infty] \lambda^{\frac{2s+1}{2}} dp_1\;\;
s=0,1,\dots,.
\end{equation}
Then, the equations
\begin{equation}
\label{whitham} u^i_{t_s}=J^{ij}_0\displaystyle\frac{\delta
h_{s+1}}{\delta u^j}=J^{ij}_1\displaystyle\frac{\delta
h_{s}}{\delta u^j}=v_{(s)}^i u^i_x, \;\;s=0,1,\dots,
\end{equation}
corresponds to the KdV-Whitham hierarchy with $t_0=x$ and $t_1=t$ \cite{FFM},
\cite{Krichever}.

The metrics $ \displaystyle \frac{g^0_{ii}}{(u^i)^s}$,
$s=2,3,\dots$, are not flat and the non-zero elements of the
associated curvature tensor $R^{ij}_{ij}(\vu,s)$ are
\begin{equation}
\label{curvature0} R^{ij}_{ij}(\vu,s)=-\displaystyle\frac{1}{2
\sqrt{g_{ii}^0g_{jj}^0}}\left(
\sum_{k=1}^{2g+1}(u^k)^{s}\partial_{u^k}W(P_i,P_j)+
\displaystyle\frac{s}{2}((u^i)^{s-1}+(u_j)^{s-1}))W(P_i,P_j)\right).
\end{equation}
Formulas (\ref{curvature0}) hold true also for $s=0,1$, where the
r.h.s. identically vanishes (a proof may be found in
\cite{Du1},\cite{KK}).

In the next lemma we evaluate the curvature tensor (\ref{curvature0}) for $s=2,3$.
\begin{lemma}
The metrics
\begin{equation}
\label{metric2}
g^2_{ii}=\displaystyle\frac{g^0_{ii}}{(u^i)^2},\quad
g^3_{ii}=\displaystyle\frac{g^0_{ii}}{(u^i)^3}
\end{equation}
are constant curvature and conformally flat respectively. The
nonzero elements of the curvature tensor (\ref{curvature0}) take
the form
\begin{equation}
\label{curvature00} \quad
R^{ij}_{ij}(\vu,s=2)=-\displaystyle\frac{1}{2},\quad \quad
R^{ij}_{ij}(\vu,s=3)= -
\displaystyle\frac{3}{2}\left(\displaystyle\frac{dp_2(P_i)}{dp_1(P_i)}+
\displaystyle\frac{dp_2(P_j)}{dp_1(P_j)}\right),
\end{equation}
with $dp_{1,2}$ as in (\ref{dp}).
\end{lemma}
{\it Proof}.
Using the fact that (\ref{curvature0}) vanishes for $s=0,1$ we obtain
\begin{align*}
\sum_{k=1}^{2g+1}(u^k)^{s}\partial_{u^k}W(P_i,P_j)=&\sum_{k\neq
i,j}(u^k)^s \partial_kW(P_i,P_j)+
\displaystyle\frac{(u^i)^s}{u^i-u^j}\sum_{k\neq i,j}(u^j-u^k)\partial_kW(P_i,P_j)\\
-& \displaystyle\frac{(u^j)^s}{u^i-u^j}\sum_{k\neq
i,j}(u^i-u^k)\partial_kW(P_i,P_j)
+\displaystyle\frac{(u^i)^s-(u^j)^s}{u^i-u^j}W(P_i,P_j)
\end{align*}
Using (\ref{W-variation}) and the residue theorem we re-write the above in the form
\begin{align*}
\sum_{k=1}^{2g+1}(u^k)^{s}\partial_{u^k}W(P_i,P_j)=&-
\res[P=P_i,P_j,\infty]\lb(P)^s\displaystyle\frac{W(P_i,P)W(P_j,P)}{d\lb(P)}\\
&-\displaystyle\frac{(u^i)^s}{u^i-u^j}\res[P=P_i,\infty](\lb(P_j)-\lb(P))
\displaystyle\frac{W(P_i,P)W(P_j,P)}{d\lb(P)}\\
&+\displaystyle\frac{(u^j)^s}{u^i-u^j}\res[P=P_j,\infty](\lb(P_i)-\lb(P))
\displaystyle\frac{W(P_i,P)W(P_j,P)}{d\lb(P)}
\end{align*}
The last two terms in the r.h.s. of the above expression are
holomorphic at infinity so that
\begin{align*}
&\sum_{k=1}^{2g+1}(u^k)^{s}\partial_{u^k}W(P_i,P_j))+
\displaystyle\frac{s}{2}((u^i)^{s-1}+(u_j)^{s-1}))W(P_i,P_j)\\
&=-
\res[P=\infty]\lb(P)^s\displaystyle\frac{W(P_i,P)W(P_j,P)}{d\lb(P)}\\
&-(u^i)^s\res[P=P_i]\left[\left((1+\displaystyle\frac{\lb(P_j)-\lb(P)}{u^i-u^j}\right)
\displaystyle\frac{W(P_i,P)W(P_j,P)}{d\lb(P)}\right]\\
&-(u^j)^s\res[P=P_j]\left[\left((1+\displaystyle\frac{\lb(P_i)-\lb(P)}{u^j-u^i}\right)
\displaystyle\frac{W(P_i,P)W(P_j,P)}{d\lb(P)}\right]
\end{align*}
For $s=0,1$  the first term in the r.h.s of the above expression
vanishes because it is holomorphic at infinity. Since  for $s=0,1$
the curvature tensor $R^{ij}_{ij}(\vu,s=0,1)$ is equal to zero, it
follows that the last two terms of the above expression are
identically zero. So we conclude that the curvature tensor takes
the form
\begin{align}
\label{curvature0b} R^{ij}_{ij}(\vu,s)&=-\displaystyle\frac{
\res[P=\infty]\left[\lb(P)^s\displaystyle
\frac{W(P_i,P)W(P_j,P)}{d\lb(P)}\right]}{dp_i(u^i)dp_1(u^j)}
\\
&\quad=\left\{\begin{array}{ll}
-\displaystyle\frac{1}{2}& \text{for }s=2\\
 - \displaystyle\frac{3}{2}\left(\displaystyle\frac{dp_2(P_i)}{dp_1(P_i)}+
\displaystyle\frac{dp_2(P_j)}{dp_1(P_j)}\right)& \text{for }s=3.
\end{array}\right.
\end{align}
The lemma is proved.

\smallskip

As a first application of the theorems in Section~\ref{sec5} on sufficient conditions
for a reciprocal metric to be flat, we consider the reciprocal
transformation of $x$, $d{\hat x} = r^a dx + A^{(a,j)} dt_j$,
where $r^a$, $a=1,\dots,g$, are Casimirs common to all the metrics
$g_{ii}^s$, $u^i_{t_j} = v_{(j)}^i u^i_x$ is the $j$--th modulated
flow of the KdV hierarchy (\ref{whitham}) and $A^{(a,j)}$ makes the transformation
closed. Then the following results can be obtained in a
straightforward way applying theorem~\ref{equiv}.
\begin{theorem}
Let $g_{ii}^s$, $s=0,1,2,3$ be the metrics defined in
(\ref{metric}) and (\ref{metric2}). Then the reciprocal metrics
\begin{equation}
\label{metricC}
\frac{g_{ii}^s}{(r^a)^2},\;\;a=1,\dots,g,\; s=0,1,2,3,
\end{equation}
where $r^a$, $a=1,\dots,g,$ are the Casimirs defined in
(\ref{flatc1}), are flat compatible diagonal metrics.
\end{theorem}
{\it Proof}. In order to prove that the metrics (\ref{metricC})
are flat,  it is sufficient to verify the condition (\ref{nablab}).
For $s=0$ and $s=1$, the quantity $T$ defined in remark (\ref{rem1}) is equal
to zero and the condition
(\ref{nablab}) takes the form
\[
\sum_{i=1}^{2g+1}\frac{(\partial_i r^a)^2(u^i)^s}{g^0_{ii}}=0, \quad
s=0,1,\;\;,\,a=1,\dots,g.
\]
In the following we prove the above relation.
Using the variational formula (\ref{W-variation}) we obtain
\[
\partial_i r^a=\frac{1}{2}dp(P_i)\omega_a(P_i)
\]
so that
\begin{equation}
\begin{split}
\label{exp1} \sum_{i=1}^{2g+1}\frac{(\partial_i
r^a)^2(u^i)^s}{g^0_{ii}}&=\displaystyle\frac{1}{2}
\sum_{i=1}^{2g+1}(u^i)^s \omega_a(P_i)^2\\&=\sum_{i=1}^{2g+1}
\res[P=P_i]\frac{\lb^s(\omega_a(\lb))^2}{d\lb}=0, \quad
s=0,1,\;\;a=1,\dots,g,
\end{split}
\end{equation}
because  $\displaystyle\frac{\lb^s(\omega(\lb))^2}{d\lb}$,
$s=0,1$, is a differential with simple poles at the branch points
$P_i$ and regular at infinity and  therefore, the sum of all its
residue is equal to zero. For the metric $g_{ii}^2$ the condition
(\ref{nablab}) takes the form
\begin{equation}\label{s2}
\sum_{i=1}^{2g+1}\frac{(\partial_i
r^a)^2(u^i)^2}{g^0_{ii}}-\displaystyle\frac{1}{2}(r^a)^2=0, \quad
s=0,1,\;\;a=1,\dots,g.
\end{equation}
To prove the above relation we use (\ref{exp1}) and then evaluate
the residue at infinity  obtaining
\[
\sum_{i=1}^{2g+1}\frac{(\partial_i r^a)^2(u^i)^2}{g^0_{ii}}=
\sum_{i=1}^{2g+1}\res[P=P_i]\frac{\lb^2(\omega_a(\lb))^2}{d\lb}
=-\res[\lambda=\infty]\frac{\lb^2(\omega_a(\lb))^2}{d\lb}=
\displaystyle\frac{1}{2}(r^a)^2
\]
because of the Riemann bilinear relations
\begin{equation}
\label{br}
r^a=\res[\lambda=\infty]p_1(\lb)\omega_a(\lb)=
\res[\lambda=\infty]\displaystyle\sqrt{\lb}\omega_a(\lb).
\end{equation}
For the metric $g_{ii}^3$ the condition (\ref{nablab}) takes the
form
\[
\sum_{i=1}^{2g+1}\frac{(\partial_i
r^a)^2(u^i)^3}{g^0_{ii}}-\displaystyle\frac{3}{4\pi i
}r^a\oint_{b_a}dp_2=0, \quad a=1,\dots,g,
\]
since $T=-\displaystyle\frac{3}{4\pi i }\oint_{b_a}dp_2$.
To prove the above relation we use (\ref{exp1}) and then evaluate
the residue at infinity  obtaining
\[
\sum_{i=1}^{2g+1}\frac{(\partial_i r^a)^2(u^i)^3}{g^0_{ii}}=
-\res[\lambda=\infty]\frac{\lb^3(\omega_a(\lb))^2}{d\lb}=
\displaystyle\frac{3}{4\pi i }r^a\oint_{b_a}dp_2,
\]
because of the Riemann bilinear relations (\ref{br}) and
\[
\displaystyle\frac{1}{2\pi
i}\oint_{b_a}dp_2=\res[\lambda=\infty]p_2(\lb)\omega_a(\lb)=
\displaystyle \frac{1}{3}\res[\lambda=\infty]\displaystyle
\lb^{\frac{3}{2}}\omega_a(\lb).
\]
The theorem is proved.
\hspace{.5 truecm} $\square$

\smallskip

As a second application of the theorems on sufficient conditions
for the reciprocal metric to be flat, we consider the reciprocal
transformation of $x$, $d{\hat x} = p_1(0) dx + A^{(p,l)} dt_l$,
where $p_1(0)$ is the Casimir associated to $g_{ii}^1(\vu)$ which
generates the modulated first negative KdV flow (index $l=-1$ in
the transformation). In the case of genus $g=1$,  we
showed in \cite{AG} that this reciprocal transformation relates the modulated
first negative KdV flow and the modulated Camassa--Holm equations.
Next theorem generalizes such relation to any genus and can be
obtained in a straightforward way applying theorem~\ref{equiv}.

\begin{theorem}\label{teo8.4}
Let $g_{ii}^s$, $s=2,3$  be the metrics defined in
(\ref{metric2}). Then the reciprocal metrics
\begin{equation}
\label{metricD}
\frac{g_{ii}^s}{p_1(0)^2},\; s=2,3,
\end{equation}
where $p_1(0)$ is the Casimir for $g_{ii}^1$ defined in
(\ref{flatc2}) are flat compatible diagonal metrics.
\end{theorem}
{\it Proof}. In order to prove that the metrics (\ref{metricD})
are flat, it is sufficient to verify the condition (\ref{nablab}).
For the metric $g_{ii}^2$ the condition (\ref{nablab}) takes the
form
\begin{equation}
\label{CH0} \sum_{i=1}^{2g+1}\frac{(\partial_i
p_1(0))^2(u^i)^2}{g^0_{ii}}-\displaystyle\frac{1}{2(p_1(0))^2}=kp_1(0),
\end{equation}
where $k$ is a constant. To prove the above relation we first
observe that
\[
\res[\lambda=\infty]\lb^{\frac{1}{2}}\Lambda_0(\lb)=-2p_1(0),
\]
where $\Lambda_0(\lb)$ is a normalized third kind differential
with simple pole in $(0,\pm\sqrt{\prod_{k=1}^{2g+1}(-u^k)})$ with
residues $\pm 1$ respectively. Applying (\ref{W-variation}) we
deduce
\[
\partial_i p_1(0)=\displaystyle\frac{1}{4}dp_1(u^i)\Lambda_0(u^i).
\]

Then we reduce the sum in (\ref{CH0}) to the evaluation of  a
residue
\begin{equation}
\label{CH1} \sum_{i=1}^{2g+1}\frac{(\partial_i
p_1(0))^2(u^i)^2}{g^0_{ii}}=
\sum_{i=1}^{2g+1}\res[P=P_i]\frac{\lb^2(\Lambda_0(\lb))^2}{4d\lb}=-
\res[P=\infty]\frac{\lb^2(\Lambda_0(\lb))^2}{4d\lb}
=\displaystyle\frac{1}{2}(p_1(0))^2.
\end{equation}
From the above relation we conclude that (\ref{CH0}) is satisfied
with $k=0$.

For the metric $g_{ii}^3$ the condition (\ref{nablab}) takes the
form
\begin{equation}
\label{CH2}
\sum_{i=1}^{2g+1}\frac{(\partial_i p_1(0))^2(u^i)^3}{g^0_{ii}}-3p_1(0) p_2(0)=0,
\end{equation}
because $T=-\displaystyle\frac{3}{2}p_2(0)$. To prove the above
relation we use (\ref{CH1})  obtaining
\[
\sum_{i=1}^{2g+1}\frac{(\partial_i p_1(0))^2(u^i)^3}{g^0_{ii}}
=-\res[P=\infty]\frac{\lb^3(\Lambda_0(\lb))^2}{4d\lb}=
3p_1(0) p_2(0).
\]
because
$\res[\lambda=\infty]\lb^{\frac{3}{2}}\Lambda_0(\lb)=-6p_2(0)$.
The above relation shows the validity of (\ref{CH2}). \hspace{.5
truecm} $\square$

As a third example we consider the Casimir $h_0$ that generates
the positive KdV modulated flows ($s=1$ in (\ref{whitham})).
\begin{lemma}
The metric
\begin{equation}
\label{q1} \displaystyle\frac{g_{ii}^0}{h_0^2 u^i}
\end{equation}
is flat with  $h_0$ the Casimir for $g_{ii}^0$ defined in
(\ref{flatc1}). The metrics
\begin{equation}
\label{q2} \displaystyle\frac{g_{ii}^0}{h_0^2},\quad
\displaystyle\frac{g_{ii}^0}{(h_0u^i)^2},
\end{equation}
are respectively  constant curvature and conformally flat.
\end{lemma}
{\it Proof}. To prove the lemma  we need the relation
\begin{align}
\nonumber
\sum_{i=1}^{2g+1}\frac{(\partial_i h_0)^2(u^i)^s}{g^0_{ii}}&=
\sum_{i=1}^{2g+1}(dp_1(u^i))^2(u^i)^s=\sum_{i=1}^{2g+1}
\res[P=P_i]\displaystyle\frac{\lb^s dp_i(\lb)^2}{d\lb}\\
\label{pq1} &=- \res[\infty]\displaystyle\frac{\lb^s
dp_i(\lb)^2}{d\lb}= \left\{\begin{array}{ll}
-2, &\text{for  } s=0\\
2h_0, &\text{for  } s=1\\
\displaystyle\frac{1}{2}h_0^2+2h_1, &\text{for  } s=2
\end{array}\right.
\end{align}

where $h_1$  has been defined in (\ref{kdvh}).

In order to prove that the  metric (\ref{q1}) is  flat, it is
sufficient to verify the condition (\ref{nablab}), where $h_0$ is
a density of momentum for the metric $g_{ii}^0/u^i$, that is there
exists a constant $k$ such that
\[
\sum_{i=1}^{2g+1}\frac{(\partial_i h_0)^2u^i}{g^0_{ii}}=kh_0
\]
and comparing with (\ref{pq1}) we find $k=2$.

The relation (\ref{pq1}) immediately implies that the  metrics in
(\ref{q2}) are respectively with constant curvature $-2$ and
conformally flat with affinors $\tilde{v}^i$
\[
\tilde{v}^i=h_0\displaystyle\frac{\partial_i h_1}{\partial_i
h_0}-h_1.
\]
$\quad \square$
\smallskip

As a final example we consider a reciprocal transformation of $x$ and $t$ of the form
\begin{eqnarray}
\left\{
\begin{array}{ll}
d\hat{x}&=r^adx+A^a dt\\
d\hat{t}&=h_0 dx+M dt
\end{array}\right.
\end{eqnarray}
where $h_0$ and $r^a$ are the Casimirs defined in (\ref{flatc1}) for the metric
$g_{ii}^0$ and $A^a$ and $M$ are the terms which make the above two  1-forms closed
with respect to the first Whitham-KdV flow, that is
\[
u^i_t=v_{(1)}^i u_x^i=(J^0)^{ij}\displaystyle\frac{\delta h_2}{\delta
u^j}, i=1,\dots,2g+1.
\]
where the hamiltonian density $h_2$ is defined in (\ref{kdvh}).

Let $\hat{g}_{ii}^0$ be the transformed metric of $g_{ii}^0$ given
by the relation (\ref{tramet})
\begin{equation}
\label{mxt}
{\hat g}^0_{ii}=
\left(\frac{M-h_0 v^i}{r^a M-A^a h_0}
\right)^2{ g}^0_{ii}.
\end{equation}
\begin{theorem}
The reciprocal metrics
\[
\displaystyle\frac{{\hat
g}^0_{ii}}{u^i},\;\;\displaystyle\frac{{\hat g}^0_{ii}}{(u^i)^2}
\]
with ${\hat g}^0_{ii}$ defined in (\ref{mxt}) form  a flat pencil of metrics.
\end{theorem}

To prove the statement we apply theorem~\ref{theo7.2} with $B=r^a$ and $N=h_0$.

(\ref{exp1}) with $s=1$  gives  $b=0$ in theorem~\ref{theo7.2} and
(\ref{pq1})  for $s=1 $ is equivalent to the flatness condition
(\ref{flatc1}) of   theorem~\ref{theo7.2}, with $\nu_3=0, \nu_4=1$
and we conclude that the  metric $\displaystyle\frac{{\hat
g}^0_{ii}(\vu)}{u^i}$ is flat.

For the second metric, similarly, (\ref{s2}) gives  $b=0$ in
theorem~\ref{theo7.2} and (\ref{pq1}) for $s=2 $ is equivalent to
the flatness condition (\ref{flatc1}) of theorem~\ref{theo7.2},
with $\nu_3=1, \nu_4=0$ and we conclude that the  metric
$\displaystyle\frac{{\hat g}^0_{ii}(\vu)}{(u^i)^2}$ is flat.

\vskip 1cm {\bf Acknowledgments} This work has been partially
supported by the European Science Foundation Programme MISGAM
(Method of Integrable Systems, Geometry and Applied Mathematics)
the RTN ENIGMA and PRIN2006 ''Metodi geometrici nella teoria delle
onde non lineari ed applicazioni''.

\end{document}